\documentclass[fleqn,10pt]{wlscirep}
\usepackage[utf8]{inputenc}
\usepackage[T1]{fontenc}
\usepackage{lineno}
\usepackage[caption=false,font=normalsize,labelfont=bf,textfont=sf]{subfig}

\title{GazeBaseVR, a large-scale, longitudinal, binocular eye-tracking dataset collected in virtual reality}

\author[1,$\dag$,*]{Dillon Lohr}
\author[1,$\dag$]{Samantha Aziz}
\author[1]{Lee Friedman}
\author[1]{Oleg V Komogortsev}
\affil[1]{Texas State University, Department of Computer Science, San Marcos, TX 78666, USA}

\affil[*]{corresponding author: Dillon Lohr (djl70@txstate.edu)}

\affil[$\dag$]{these authors contributed equally to this work}

\begin{abstract}

We present GazeBaseVR, a large-scale, longitudinal, binocular eye-tracking~(ET) dataset collected at 250~Hz with an ET-enabled virtual-reality~(VR) headset.
GazeBaseVR comprises 5,020~binocular recordings from a diverse population of 407~college-aged participants.
Participants were recorded up to six times each over a 26-month period, each time performing a series of five different ET tasks: (1) a vergence task, (2) a horizontal smooth pursuit task, (3) a video-viewing task, (4) a self-paced reading task, and (5) a random oblique saccade task.
Many of these participants have also been recorded for two previously published datasets with different ET devices, and some participants were recorded before and after COVID-19 infection and recovery.
GazeBaseVR is suitable for a wide range of research on ET data in VR devices, especially eye movement biometrics due to its large population and longitudinal nature.
In addition to ET data, additional participant details are provided to enable further research on topics such as fairness.
\end{abstract}
\begin{document}

\flushbottom
\maketitle

\thispagestyle{empty}


\section*{Background \& Summary}


Eye-tracking~(ET) sensors are becoming increasingly prevalent in modern virtual- and augmented-reality~(VR/AR) devices such as the Vive Pro Eye,~\cite{ViveProEye} HoloLens~2,~\cite{HL2} and Magic Leap~2.~\cite{ML2}
The presence of these ET sensors is motivated in large part to enable foveated rendering techniques~\cite{guenter_2012_foveated3dgraphics} which offer a significant reduction in overall power consumption without a noticeable impact on visual quality.
Such power savings could lead to higher resolution displays in tethered devices or a longer battery life in untethered devices.
In addition to foveated rendering, ET also enables a multitude of applications including (continuous) user authentication,~\cite{lohr_2022_ekyttowardviable,zhang_2018_continuousauthenticationeye} health monitoring,~\cite{vidal_2012_wearableeyetracking} novel display technologies,~\cite{konrad_2019_gazecontingentocular} usability assessment,~\cite{poole_2006_eyetrackingusability} direct gaze interaction,~\cite{sibert_2000_evaluationeyegaze} and more.

Research on these applications is heavily dependent on the availability of large-scale datasets.
Eye movement biometrics~(EMB),~\cite{kasprowski_2004_eyemovementsbiometrics} especially, requires large, longitudinal datasets with hundreds of unique identities and varied eye movement behaviors to train state-of-the-art deep learning models.~\cite{lohr_2022_ekyttowardviable,makowski_2021_deepeyedentificationlive}
One of the most suitable datasets for EMB is GazeBase,~\cite{griffith_2021_gazebaselargescale} a dataset of high-quality monocular (left eye only) ET signals recorded at 1000~Hz over a 37-month period from a population of 322~college-aged participants.
However, at the time of writing, there is no similar, large-scale, longitudinal dataset collected with a modern VR/AR device, making it difficult to train an EMB model for such devices.

\begin{figure}[ht]
    \centering
    \subfloat[][]{
        \label{fig:setup}
        \includegraphics[height=2.5in,trim={0 2in 0 5in},clip]{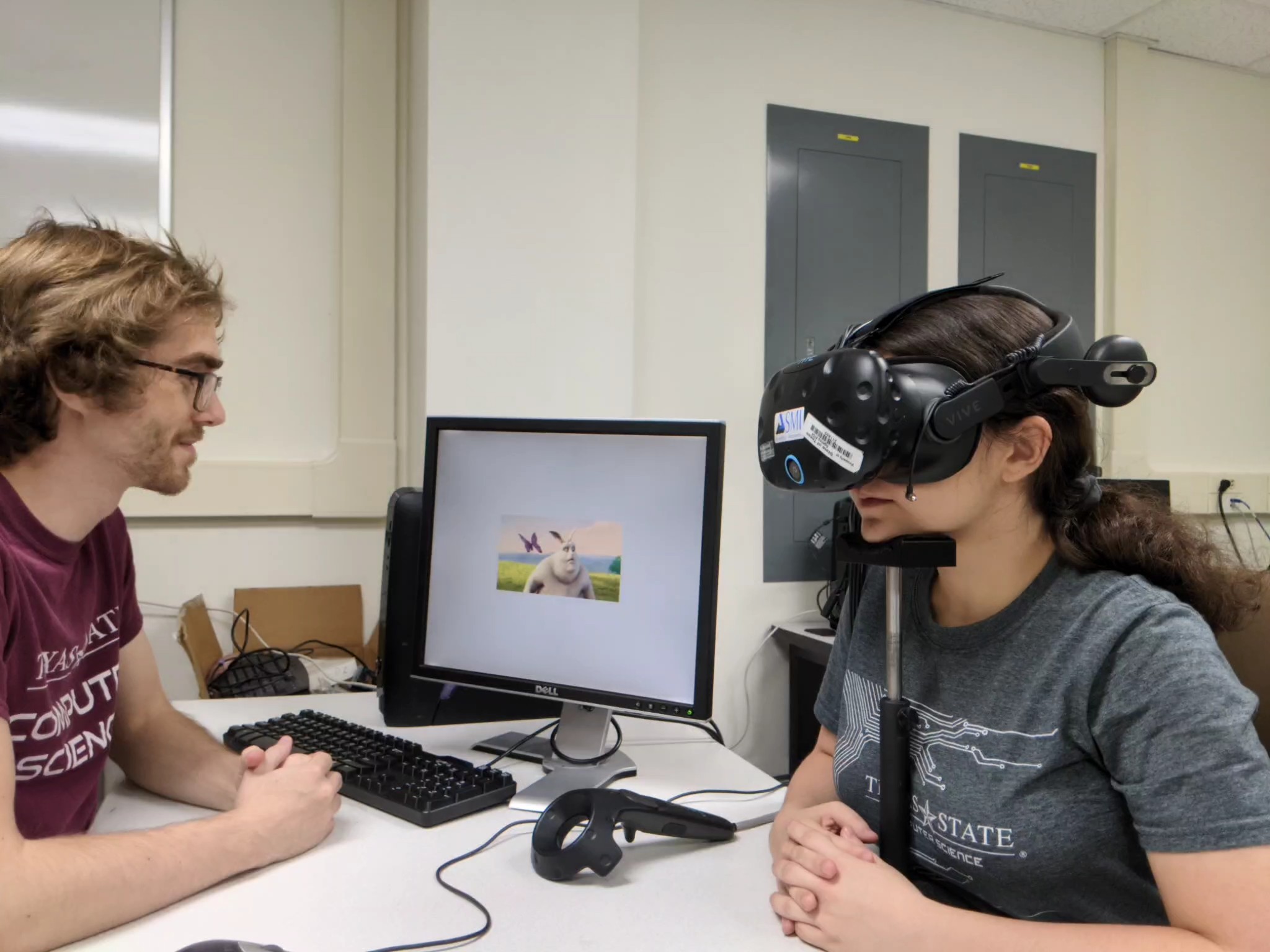}
    } \\
    \subfloat[][]{
        \label{fig:task_vergence}
        \includegraphics[scale=1.5]{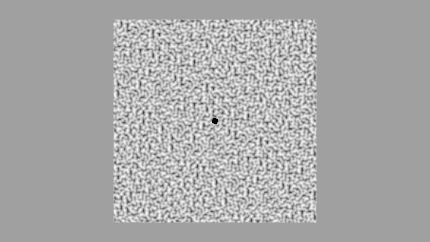}
    }
    \subfloat[][]{
        \label{fig:task_video}
        \includegraphics[scale=1.5]{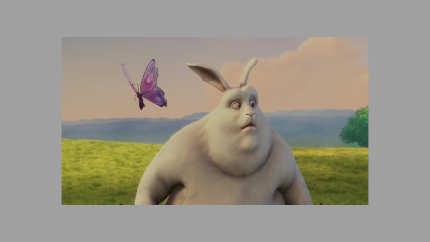}
    } \\
    \subfloat[][]{
        \label{fig:task_reading}
        \includegraphics[scale=1.5]{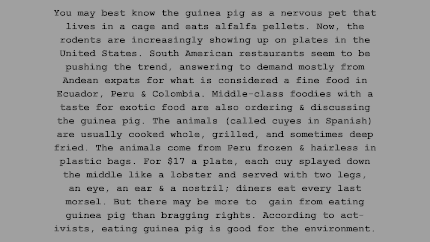}
    }
    \subfloat[][]{
        \label{fig:task_dot}
        \includegraphics[scale=1.5]{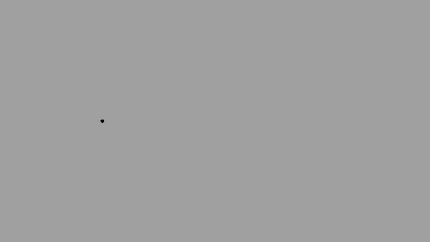}
    }
    \subfloat[][]{
        \label{fig:task_blink}
        \includegraphics[scale=1.5]{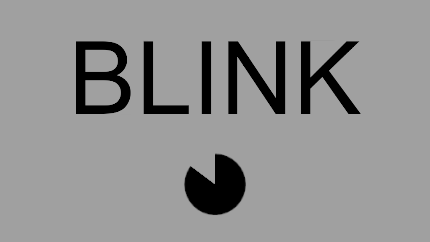}
    } \\
    \subfloat[][]{
        \label{fig:dates}
        \includegraphics[width=0.95\linewidth]{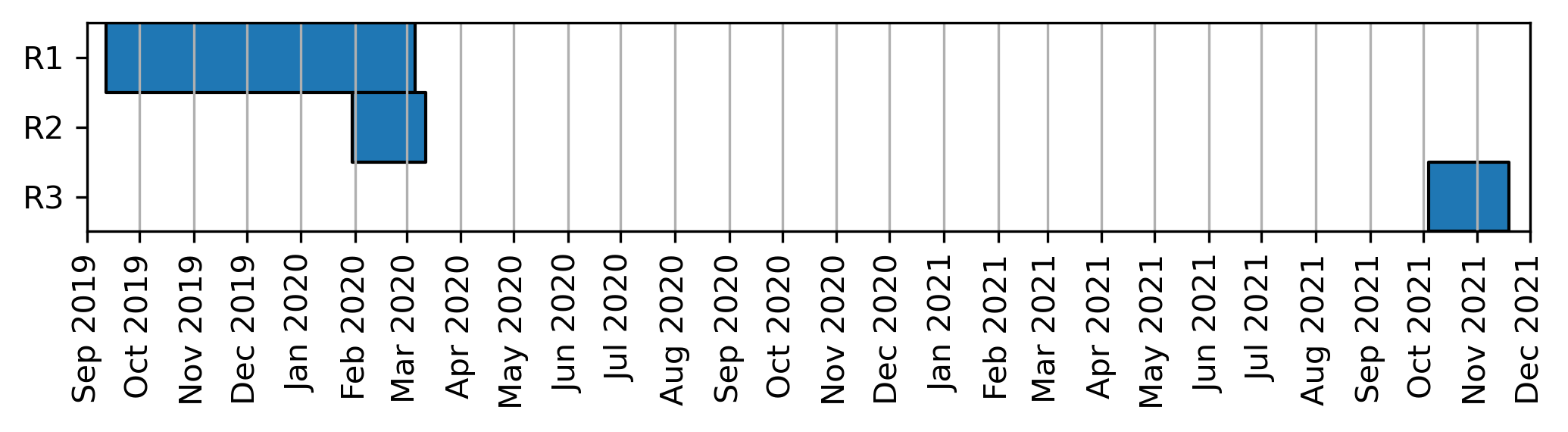}
    }
    \caption{Summary of the GazeBaseVR dataset collection. (\textbf{A}) An illustration of the experimental setup. (\textbf{B}) The stimulus used for the vergence task. (\textbf{C}) A frame from one of the video clips used for the video task. (\textbf{D}) One of the text excerpts used for the reading task. (\textbf{E}) The dot stimulus used for the smooth pursuit and random saccade tasks. (\textbf{F}) The text and timer displayed during the ``blink period'' prior to each task. (\textbf{G}) The time periods during which each recording round took place.}
    \label{fig:overview}
\end{figure}

The present work introduces GazeBaseVR, a GazeBase-inspired dataset collected with an ET-enabled VR headset.
GazeBaseVR contains binocular ET signals recorded at 250~Hz over a 26-month period from a diverse population of 407~college-aged participants.
A summary of the GazeBaseVR dataset collection is presented in Figure~\ref{fig:overview}.
One particularly noteworthy component of GazeBaseVR is a task that elicits vergence eye movements which are underrepresented in public ET datasets.
Additionally, since some participants were recorded before and after COVID-19 infection and recovery, GazeBaseVR may offer a rare opportunity to study potential lasting effects~\cite{cena_2022_eyemovementalterations} of COVID-19 on eye movement behavior.
A subset of this dataset was described and used in a brief prior study,~\cite{lohr_2020_eyemovementbiometrics} but this is the first release of the full dataset.

Further, some of the participants from GazeBaseVR were also later recorded for two other public datasets: SynchronEyes~\cite{aziz_2022_synchroneyesnovelpaired} and the HoloLens~2 ET dataset by Aziz and Komogortsev.~\cite{aziz_2022_assessmenteyetracking}
This overlap in populations may enable research on the generalizability of EMB models across several different ET devices, among other potential applications.
While many prior studies have also recorded a set of participants with multiple ET devices,~\cite{ehinger_2019_newcomprehensiveeye,holmqvist_2017_commonpredictorsaccuracy,holmqvist_2022_smallheadmovements,spitzer_2022_testbatterycompare} existing datasets tend to either not be publicly available, not contain enough unique identities for a robust analysis, or not contain sufficiently varied eye movement behaviors for applications such as EMB.

\section*{Methods}






 

\subsection*{Participants}
\begin{figure}[ht]
    \centering
    \includegraphics{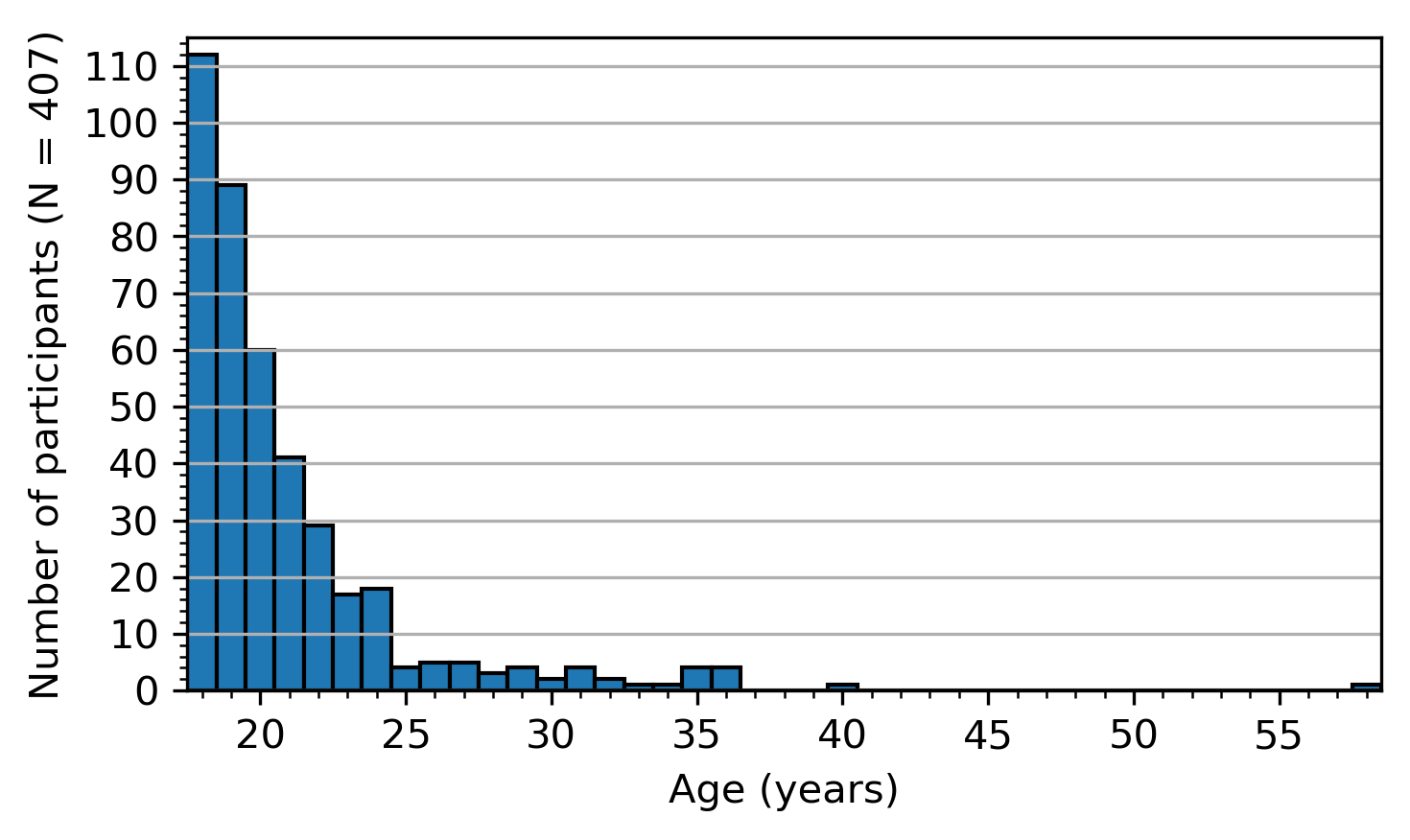}
    \caption{Distribution of participants' ages at enrollment time (i.e., the first session of the first recording round).}
    \label{fig:ages}
\end{figure}

A total of 465~individuals originally participated in the study, but 58~participants were excluded for various reasons (e.g., could not be tracked/calibrated, experienced motion sickness, could not finish within the allotted time of 1~hour, excessive (over 50\%) data loss in one or more recordings).
At the time of Round~1, 188~participants self-identified as male, 216~self-identified as female, and 3~self-identified as neither male nor female.
See Table~\ref{tab:race} for race/ethnicity statistics and Figure~\ref{fig:ages} for the distribution of participants' ages at enrollment time.
A total of 3~recording rounds took place over a period of 26~months (see Table~\ref{tab:dates} for date ranges and population sizes), with each round comprising 2~recording sessions separated by approximately 30~minutes.
Round~2 began alongside a continuation of Round~1 in the beginning of the Spring 2020 semester, but recordings were prematurely halted due to health concerns at the start of the global COVID-19 pandemic.
Recordings later resumed with Round~3, throughout which the laboratory personnel and participants all wore face masks to reduce health risks, and extra care was taken to disinfect all the equipment after each set of recordings.
There are fewer participants in Round~2 than the other rounds because it was prematurely halted due to COVID-19, so participants for Round~3 were recruited from the Round~1 population.
The participants in Rounds~2 and 3 are both subsets of the population from Round~1, but a participant in Round~2 may not be present in Round~3 and vice versa.

Participants were recruited from the undergraduate student population at Texas State University in San Marcos, TX, USA.
All participants were screened to ensure they had no history of epilepsy or seizures, and they all provided informed, written consent to participate in the study and to have their anonymized data shared with the broader research community following a protocol approved by the Institutional Review Board at Texas State University.
Participants were compensated with extra course credit for participation in Round~1, \$20 in Round~2, and \$40 in Round~3.

\subsection*{Data acquisition overview}
ET data are recorded with SensoMotoric Instrument's~(SMI's) tethered ET VR head-mounted display based on the HTC Vive (hereon called the ET-HMD).
The ET-HMD tracks both eyes at a nominal sampling rate of 250~Hz with a manufacturer-reported typical spatial accuracy of $0.2$~degrees of visual angle~(dva).
The experiments were designed in Unity~2018.3.11f1 using the C\# programming language.

To facilitate a better and more comfortable headset fit, the stock head strap was replaced by the HTC Vive Deluxe Audio Strap, but no audio was ever played during the experiments.
Additionally, the stock 14~mm foam face cushion was replaced with a 6~mm polyurethane leather face cushion to increase the field of view within the headset and make it easier to clean.

The view in the headset is fixed during each task so that, regardless of any head movement, each stimulus maintains the correct position relative to the headset.
A participant puts on the headset, adjusts its fit for comfort and image clarity, and rests his/her head on a chin rest to minimize head movements.
Although the view in the headset is fixed, it is still desirable to minimize head movements with a chin rest to reduce headset slippage, to reduce the risk of discomfort caused by the fixed view, and to reduce unintended eye movement behavior caused by the vestibulo-ocular reflex.

\subsection*{Calibration and validation}
Participants perform a manufacturer-provided calibration procedure at scheduled intervals prior to the vergence, reading, and random saccade tasks, or whenever the headset is removed for any reason.
The calibration procedure involves following a moving dot in a standard 5-point grid pattern.
Calibration is performed up to 3~times until a spatial accuracy below $1$~dva is achieved, moving on after the third attempt regardless of spatial accuracy.
Spatial accuracy is assessed after each calibration attempt with a short, custom validation procedure consisting of a 13-point grid spanning $\pm 15$~dva horizontally and $\pm 10$~dva vertically at a depth of 1~meter.

In an effort to reduce fatigue, calibration is not performed prior to every task.
This is justified by the use of a head-mounted display and a chin rest, as significant headset slippage is unlikely and high tracking accuracy can be maintained for longer periods than may be expected for non-wearable ET devices.

\subsection*{Task battery overview}
An ordered series of 5~eye-tracking tasks are performed during each recording session.
Each task is described in the following subsections in the order they occur within each session.
The task abbreviations included in the subsection titles are part of the file naming convention for GazeBaseVR.

Each task is preceded by a 3-second-long ``blink period'' during which participants are instructed to blink as needed in an effort to reduce the amount they would need to blink during the task itself.
Any eye-tracking data recorded during these blink periods is discarded.
During this period, the text ``BLINK'' appears in large, black font over a light-gray background.
The same light-gray background color is used for all tasks.
Below the text is a black radial wipe timer that participants can use to gauge how much time remains until the task begins.
Participants are instructed to try to minimize their blinks during each task and, if they need to blink, to try to blink only during periods when the visual target is stationary.
See Figure~\ref{fig:task_blink} for a visualization of the blink period.

\subsubsection*{Task 1: Vergence Task (VRG)}
This task is modeled after a study on the dynamics of vergence eye movements by Tyler et al.~\cite{tyler_2012_analysishumanvergence}
During the task, a large ($30 \times 30$~dva), square plane textured with random, gray-scale noise is displayed in the center of the user's field of view.
At the center of the plane is a small ($1$~dva diameter), black sphere on which participants are instructed to focus throughout the task.
The stimulus alternates between depths of approximately 0.4433 and 0.3543~meters, eliciting ideal vergence (left minus right) of 8 and 10 dva, respectively, assuming an interocular distance of 62~mm.
The stimulus scales in size with changes in depth to maintain a constant apparent size so that vergence eye movements are driven by image disparity alone.
Periods between depth changes are uniformly random between 2 and 3~seconds.
The task has a duration between 48 and 72~seconds and elicits a total of 12~convergent (toward the nasal bridge) and 12~divergent (away from the nasal bridge) eye movements.
See Figure~\ref{fig:task_vergence} for a visualization of the stimulus for this task.

\subsubsection*{Task 2: Smooth Pursuit Task (PUR)}
During this task, a small, black sphere ($0.5$~dva diameter at 1~meter depth) glides smoothly between the left and right edges of the viewing region ($\pm 15$~dva) to elicit horizontal smooth pursuit eye movements.
The stimulus begins at the center of the screen and, after a delay of 1.5~seconds, smoothly moves to the left edge of the viewing region at a constant speed of $5$~dva/s.
After a random delay between 1 and 1.5~seconds, it smoothly moves from the left edge to the right edge, pauses for another random delay when it reaches the right edge, and then smoothly moves back to the left edge.
We refer to this complete left-to-right-to-left movement as a ``trap,'' since when plotting the horizontal position of the stimulus versus time its shape resembles a trapezoid.
The stimulus performs as many complete traps as necessary to satisfy at least 30~seconds of movement, not including the random pauses at the left and right edges nor the time it takes to move to and from the center of the screen.
It then returns to the center of the viewing region, pauses for 1.5~seconds, and repeats the full movement pattern at a higher speed.

A total of 3~different speeds are employed during this task in a fixed order: $5$~dva/s, $10$~dva/s, and $20$~dva/s.
At $5$~dva/s, the stimulus performs 3~traps totaling 36~seconds of movement, plus an additional 6~seconds moving to and from the center of the screen.
At $10$~dva/s, the stimulus performs 5~traps totaling 30~seconds of movement, plus an additional 3~seconds moving to and from the center of the screen.
At $20$~dva/s, the stimulus performs 10~traps totaling 30~seconds of movement, plus an additional 1.5~seconds moving to and from the center of the screen.
Together with the pauses at the edges and center, the task has a total duration between 151.5 and 171~seconds.
See Figure~\ref{fig:task_dot} for a visualization of the stimulus for this task.

\subsubsection*{Task 3: Video Viewing Task (VID)}
During this task, a video ($1280 \times 720$~resolution, 30~frames per second) is displayed on a large, rectangular plane ($36 \times 21$~dva at 1~meter depth) in the center of the user's field of view.
Participants are instructed to view the video as they normally would.
The video is a clip from the 3D animated short film, Big Buck Bunny,~\cite{roosendaal_2008_bigbuckbunny} with different clips being used for each session and the same two clips being used for all recording rounds.
The first session uses the clip between timestamps 01:50--02:28 (38~seconds duration) and the second session uses the clip between timestamps 05:45--06:23 (38~seconds duration).
Each clip involves periods where one or more objects of interest are moving or stationary, eliciting a variety of eye movement behaviors.
See Figure~\ref{fig:task_video} for a visualization of the stimulus for this task.

\subsubsection*{Task 4: Reading Task (TEX)}
During this task, an excerpt (roughly 820~characters) of an article from National Geographic is displayed within a $51.2 \times 37.6$~dva viewing region at a depth of 0.6~meters in the center of the user's field of view.
The chosen text contains easily digestible, non-fiction prose.
A fixed-width font is used such that each character has a width close to $1$~dva (varying with eccentricity).
The font is black and is displayed over a light-gray background.

Participants hold an HTC Vive controller during the task and are instructed to press the rear trigger button to indicate that they have finish reading the text.
Afterward, a multiple-choice reading comprehension question is displayed in the headset and participants must select one of the four answer choices using the controller to complete the task.
Participants are informed beforehand that there will be a reading comprehension question.
The question is not intended to be difficult, and the correctness of the selected answer is irrelevant; the purpose of this question is to encourage participants to read the text closely and not merely skim through it.
The selected answer choice and any eye-tracking data recorded while answering the question are discarded.

The duration of the task depends on how quickly a participant reads through the text, ranging from 22.2 to 141.4~seconds (median~51.6, IQR~17.9).
A total of 4~unique text excerpts were used throughout data collection: one for session~1 of Round~1, one for session~2 of Round~1, one for session~1 of Rounds~2 and 3, and one for session~2 of Rounds~2 and 3.
See Figure~\ref{fig:task_reading} for a visualization of the stimulus for this task.

\subsubsection*{Task 5: Random Saccade Task (RAN)}
During this task, a small, black sphere ($0.5$~dva diameter at 1~meter depth) begins at the center of the user's field of view and jumps to uniformly random positions on the screen within $\pm 15$~dva horizontally and $\pm 10$~dva vertically.
There is a uniformly random delay between 1 and 1.5~seconds and a minimum distance of $3$~dva separating consecutive jumps.
Participants are instructed to focus on and follow the sphere with their eyes throughout the task.
A total of 79~stimulus movements (80~fixation periods) occur throughout the task, resulting in a duration between 80 and 120~seconds.
See Figure~\ref{fig:task_dot} for a visualization of the stimulus for this task.

\section*{Data Records}



GazeBaseVR is available for download on figshare~\cite{lohr_2022_gazebasevrdatarepository} under a Creative Commons Attribution 4.0 International (CC-BY~4.0) license.
In addition to the ET data, a file named \texttt{participant\_details.xlsx} is included with many self-reported details for each participant, including but not limited to age, gender, race/ethnicity, eye dominance, sleepiness on the Stanford Sleepiness Scale,~\cite{hoddes_1972_developmentusestandford} drug and alcohol use, and physical and mental health.
All recordings and the additional participant details file have been anonymized in accordance with the informed consent provided by all participants.

The ET API provided by SMI produces 3-dimensional unit vectors representing the gaze direction of each eye.
A direction vector $\mathbf{v} = [x, y, z]$ is converted to the horizontal ($\theta_H$) and vertical ($\theta_V$) components of the rotation of the eye globe in terms of dva using the equations
\begin{align}
    \theta_H &= \frac{180}{\pi} \text{atan2}\left(x, \sqrt{y^2 + z^2}\right) \\
    \theta_V &= \frac{180}{\pi} \text{atan2}\left(y, z\right),
\end{align}
where $\text{atan2}$ is the four-quadrant inverse tangent.
At each time step, a direction vector is provided for the left eye, right eye, and cyclopean eye (called the ``camera raycast'' in the ET API).
For tasks with a dot stimulus, the same equations are used to convert the position of the stimulus in world coordinates to dva relative to the cyclopean eye.

\begin{figure}[ht]
    \centering
    \includegraphics{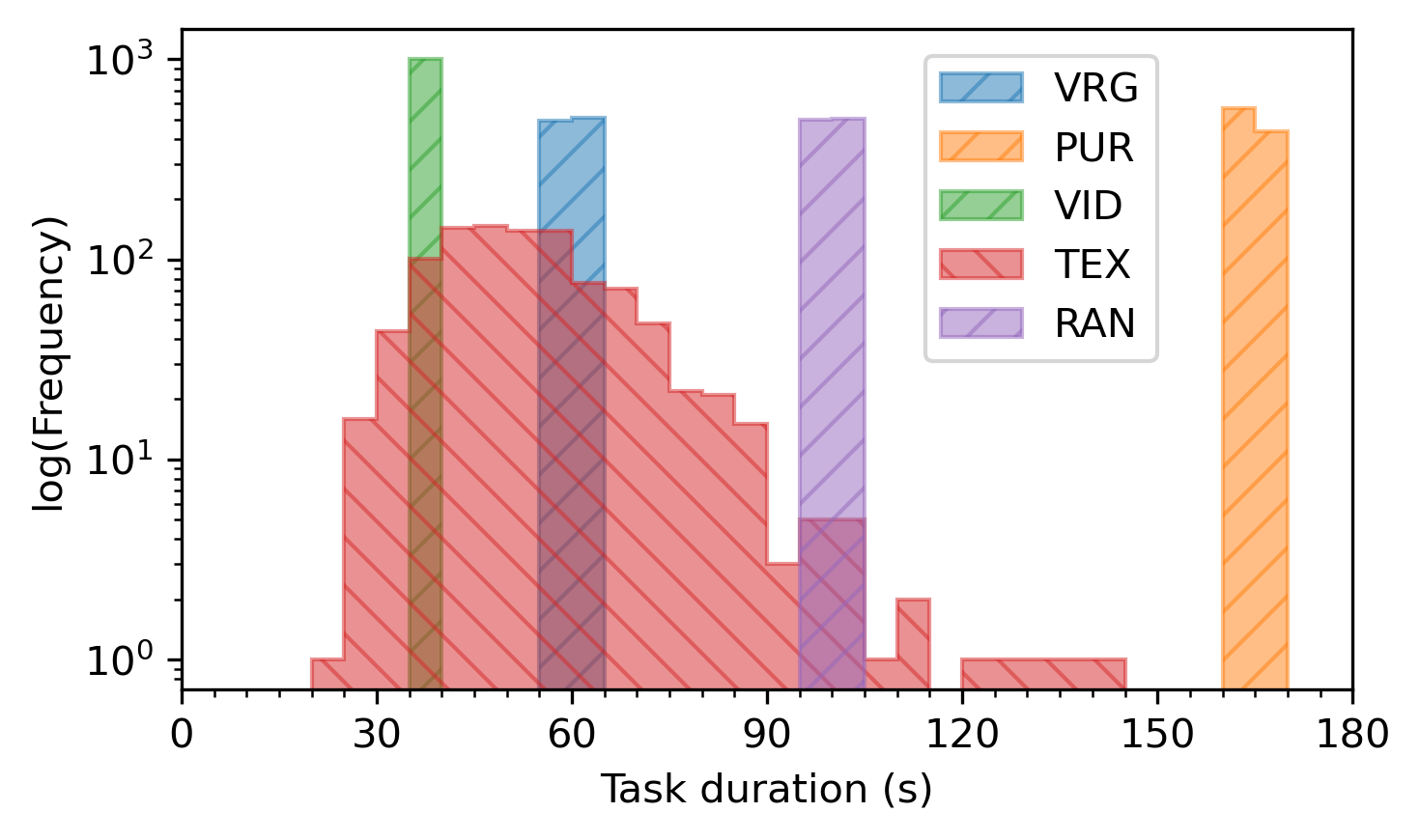}
    \caption{Distributions of the duration of each task across all recording rounds, sessions, and participants. A bin width of 5~seconds is used for each histogram.}
    \label{fig:durations}
\end{figure}

Data files are provided in CSV format inside a subdirectory named \texttt{data} following a naming convention similar to that of GazeBase: \texttt{S\_rxxx\_Sy\_z\_www.csv}.
Table~\ref{tab:filename} describes the components of the file naming convention, and Table~\ref{tab:data_records} describes the contents of each CSV file.
The distributions of recording duration grouped by task are presented in Figure~\ref{fig:durations}.

\section*{Technical Validation}


In terms of ET signal quality, the ET-HMD was one of the best ET-enabled VR headsets when it was released, boasting an impressive 250~Hz sampling rate and a manufacturer-reported typical spatial accuracy of just $0.2$~dva.
Competition at the time included devices such as the Vive Pro Eye~\cite{ViveProEye} with a sampling rate of 120~Hz and a manufacturer-reported spatial accuracy of $0.5$--$1.1$~dva, the FOVE~0~\cite{FOVE} with a sampling rate of 120~Hz and a manufacturer-reported spatial accuracy of less than $1$~dva, and the Varjo~VR-1~\cite{Varjo} with a sampling rate of 100~Hz and a manufacturer-reported spatial accuracy of less than $1$~dva.

\begin{figure}[ht]
    \centering
    \includegraphics{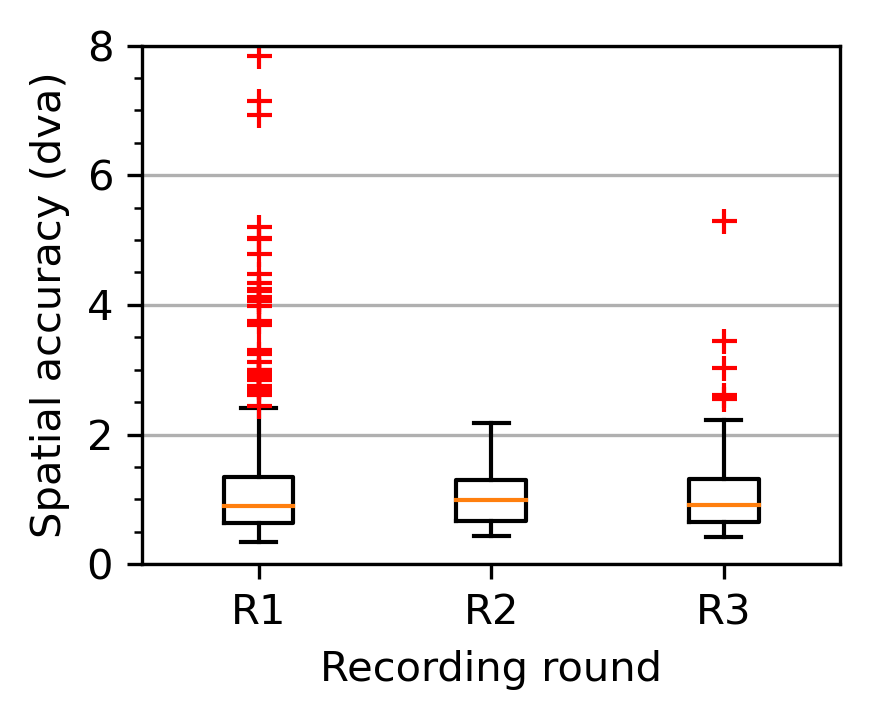}
    \caption{Rough measurements of spatial accuracy for each participant in each recording round, following the methodology of Lohr et al.~\cite{lohr_2019_evaluatingdataquality} Spatial accuracy is measured using the cyclopean gaze signal from the session~1 RAN task, averaged over the first 20~fixation periods. Saccade latency is minimized, fixation periods are identified by stimulus movement, the first 400~ms (100~samples) of each fixation period are skipped, and the next 500~ms (125~samples) are used to measure spatial accuracy. The stimulus is located at different, random positions for each fixation period, and the chosen 125-sample periods may not be ideal. But these rough measurements provide some insight into the level of signal quality contained in GazeBaseVR.}
    \label{fig:boxplot}
\end{figure}

Unlike other ET devices such as the EyeLink~1000 which provide spatial accuracy measurements during a manufacturer-provided validation procedure, we are not aware of a built-in method to quantitatively measure the spatial accuracy of the ET-HMD, at least when using the ET API within Unity.
Therefore, signal quality must be measured in a user-specified manner.
Rough measurements of the spatial accuracy of the data contained in GazeBaseVR are presented in Figure~\ref{fig:boxplot}, following the methodology from Lohr et al.~\cite{lohr_2019_evaluatingdataquality}
Based on these rough measurements, all 3~rounds have a median spatial accuracy of around $1$~dva.
Although this is significantly worse than the manufacturer-reported spatial accuracy of $0.2$~dva, it is well known that manufacturer-reported signal quality measurements are often not achievable in practice.~\cite{blignaut_2014_improvingaccuracyvideo}




\section*{Code availability}


During data collection, raw CSV files were generated from the data stream accessed with SMI's provided ET API within Unity.
These raw CSV files were later converted to the format described in Table~\ref{tab:data_records} using custom Python code.
The code used to convert the raw files to the final format, along with the code used to generate Figures~\ref{fig:dates}, \ref{fig:ages}, \ref{fig:durations}, and \ref{fig:boxplot} and the data for Tables~\ref{tab:race} and \ref{tab:dates}, is available on figshare.~\cite{lohr_2022_gazebasevrsupplementarycode}
This code was developed using Python~3.7.11 with the following main packages: numpy~1.21.6, pandas~1.3.5, openpyxl~3.0.9, and matplotlib~3.2.2.



\section*{Acknowledgements} 

This material is based upon work supported by the National Science Foundation Graduate Research Fellowship under Grant No. DGE-1840989 and DGE-1144466.
This work was also supported by the National Science Foundation under Grant No. CNS-1714623.
The authors would like to thank the numerous recording administrators who assisted with data collection.

\section*{Author contributions statement}

D.L. and S.A. wrote the manuscript, programmed the experiments, and led data quality assurance efforts; S.A. led data collection efforts and participant outreach; D.L. and L.F. led experimental design; O.K. secured funding and served as the project manager.
All authors reviewed the manuscript.

\section*{Competing interests}


The authors declare no competing interests.

\section*{Figures \& Tables}

\begin{table}[ht]
    \centering
    \begin{tabular}{lr}
        \toprule
        Race/ethnicity & Number of participants \\
        \midrule
        American Indian or Alaska Native & 0 \\
        Asian & 11 \\
        Black or African American & 41 \\
        Hispanic or Latino & 148 \\
        Native Hawaiian or Other Pacific Islander & 1 \\
        White & 140 \\
        Mixed & 62 \\
        Prefer not to answer & 4 \\
        \bottomrule
    \end{tabular}
    \caption{Self-reported race/ethnicity of the participants at the time of Round~1. Participants who self-identified as two or more options are classified as ``mixed.''}
    \label{tab:race}
\end{table}

\begin{table}[ht]
    \centering
    \begin{tabular}{llll}
        \toprule
        Recording round & Number of participants & Start date & End date \\
        \midrule
        R1 & 407 & 2019-09-12 & 2020-03-06 \\
        R2 & 35 & 2020-01-30 & 2020-03-12 \\
        R3 & 60 & 2021-10-04 & 2021-11-19 \\
        \bottomrule
    \end{tabular}
    \caption{The number of participants and dates of the first and last recordings for each recording round. Dates are given in YYYY-MM-DD format. Although Round~2 was collected concurrently with Round~1, the minimum separation between a participant being recorded for R1 and R2 was 84~days.}
    \label{tab:dates}
\end{table}

\begin{table}[ht]
    \centering
    \begin{tabular}{lll}
        \toprule
        Filename component & Description & Possible values \\
        \midrule
        \texttt{r} & Recording round & 1--3 \\
        \texttt{xxx} & Participant identifier & 001--465 \\
        \texttt{y} & Recording session & 1--2 \\
        \texttt{z} & Task number & 1--5 \\
        \texttt{www} & Task code & \texttt{VRG}, \texttt{PUR}, \texttt{VID}, \texttt{TEX}, \texttt{RAN} \\
        \bottomrule
    \end{tabular}
    \caption{Description of file naming convention: \texttt{S\_rxxx\_Sy\_z\_www}.}
    \label{tab:filename}
\end{table}

\begin{table}[ht]
    \centering
    \begin{tabular}{lll}
        \toprule
        Column header & Unit of measure & Description \\
        \midrule
        \texttt{n} & ms & timestamp of the recorded gaze sample since the beginning of the recording \\
        \texttt{x} & dva & $\theta_H$ for the cyclopean eye \\
        \texttt{y} & dva & $\theta_V$ for the cyclopean eye \\
        \texttt{lx} & dva & $\theta_H$ for the left eye \\
        \texttt{ly} & dva & $\theta_V$ for the left eye \\
        \texttt{rx} & dva & $\theta_H$ for the right eye \\
        \texttt{ry} & dva & $\theta_V$ for the right eye \\
        \texttt{xT}$^{a}$ & dva & $\theta_H$ for the stimulus, relative to the cyclopean eye \\
        \texttt{yT}$^{a}$ & dva & $\theta_V$ for the stimulus, relative to the cyclopean eye \\
        \texttt{zT}$^{b}$ & m & depth of the stimulus \\
        \bottomrule
        \multicolumn{3}{l}{$^{a}$~Only provided for tasks with a dot stimulus (VRG, PUR, and RAN). NaN for all other tasks.} \\
        \multicolumn{3}{l}{$^{b}$~Only provided for tasks with variable depth (VRG). NaN for all other tasks.}
    \end{tabular}
    \caption{Description of data format.}
    \label{tab:data_records}
\end{table}

\end{document}